\documentclass[12pt]{article}
\usepackage{graphicx}
\usepackage {amsmath}
\usepackage {amssymb}
\usepackage {longtable}
\usepackage {multirow}

\title{Equation of state of a small system with surface degrees of freedom}
\author{$^{1}$\textbf{Dmitry M. Naplekov},$^{1,2}$\textbf{Vladimir V. Yanovsky}}

\date{}

\begin{document}

\maketitle
$^{1}$\textit{Institute for Single Crystals, NAS Ukraine, 60 Nauky Ave., Kharkov, 61001, Ukraine}

$^{2}$\textit{V. N. Karazin Kharkiv National University, 4 Svobody Sq., Kharkov, 61022, Ukraine}

\begin{abstract}
We have considered a model of a small finite system with internal particles and surface degrees of freedom. All the main statistical distributions were explicitly obtained, on a pre thermodynamic limit basis. The concept of temperature or any thermodynamic equations were not used. The distribution of coordinates of a surface element allows the rigorous determination of the pressure exerted by the internal particles. In this way, we have derived the equation of state for a small system with surface. It relates the pressure to the numbers of bulk and surface degrees of freedom, their mean energies and the volume. The mean potential energy of the surface was found to be higher than the mean kinetic energy, per degree of freedom. The obtained equation of state accounts for the influence of this excessive surface energy. In the thermodynamic limit, the temperature appears and the obtained equation of state transfers to the usual ideal gas one.
\end{abstract}

\section{Introduction}

The study of small systems, whose typical sizes span from nanometers to submicron units, is a challenging frontier of modern physics. Such small-scale systems escape the concepts of classical thermodynamics \cite{eqstat-Wang,eqstat-Titu,eqstat-Zhan} that were originally developed for macroscopic systems. Classical thermodynamics deals with infinitely large systems, a condition known as the thermodynamic limit. It allows to disregard the effects associated with the surface, a finiteness of the number of degrees of freedom, fluctuations, etc. While for macroscopic systems these factors are vanishingly small, the thermodynamic limit is a good approximation. However, it is clear that as the size of the system decreases, there will be a point where this approach loses its applicability.

Typical for small systems is a substantial contribution of the surface energy, which is not proportional to the system's volume. Consequently, small systems are not extensive. Some properties of small systems are intimately related to their energy spectrum, which becomes discrete due to their finite size. The statistical properties of the phonon gas within nano-sized particles exhibit significant size dependence. The impact of fluctuations also cannot be neglected, if the number of degrees of freedom is small or the timescale is short enough. All these subjects require the development of new theoretical frameworks, relevant to the systems far from the thermodynamic limit.

Beyond the general theoretical interest, this field gains attention due to the recent experimental progress in direct manipulations and measurements of various small systems. It can be nanoparticles, thin layers, viskers or even single molecules. Many thermal, optical and other properties of these systems are found to be significantly different from those of macrosystems (see for example \cite{eqstat-Guis}). The single-molecule biological machines \cite{eqstat-Tan,eqstat-Toy}, for example, operate at energy levels of the order of thermal energy. Their operating principles are currently not fully understood. The theory level on such subjects lags significantly behind the progress made in experimental research. Even the very origin of the surface energy on an atomistic level is still far from clear \cite{eqstat-Zhao}.

Current theoretical approaches to nanoscale systems \cite{eqstat-Hil,eqstat-Schn,eqstat-Bede,eqstat-Gui,eqstat-Mig} are mostly based on the thermodynamics of macro systems, with some modifications. Other way is molecular kinetic, based on the principles of statistical mechanics. It is possible to find the exact statistical distributions for a finite system, almost without assumptions and without thermodynamic limit passage. Such distributions will depend only on mechanical quantities. This approach is implemented in this paper for the statistical description of a simple model finite system with surface degrees of freedom. First we consider the most simple case of the surface with one degree of freedom, to give the detailed explanation of our approach and obtained results. Then we extend it to the surface consisting of multiple elements.

\section{General statistical approach}

\begin{figure}
 \centering
 \includegraphics[width=12 cm]{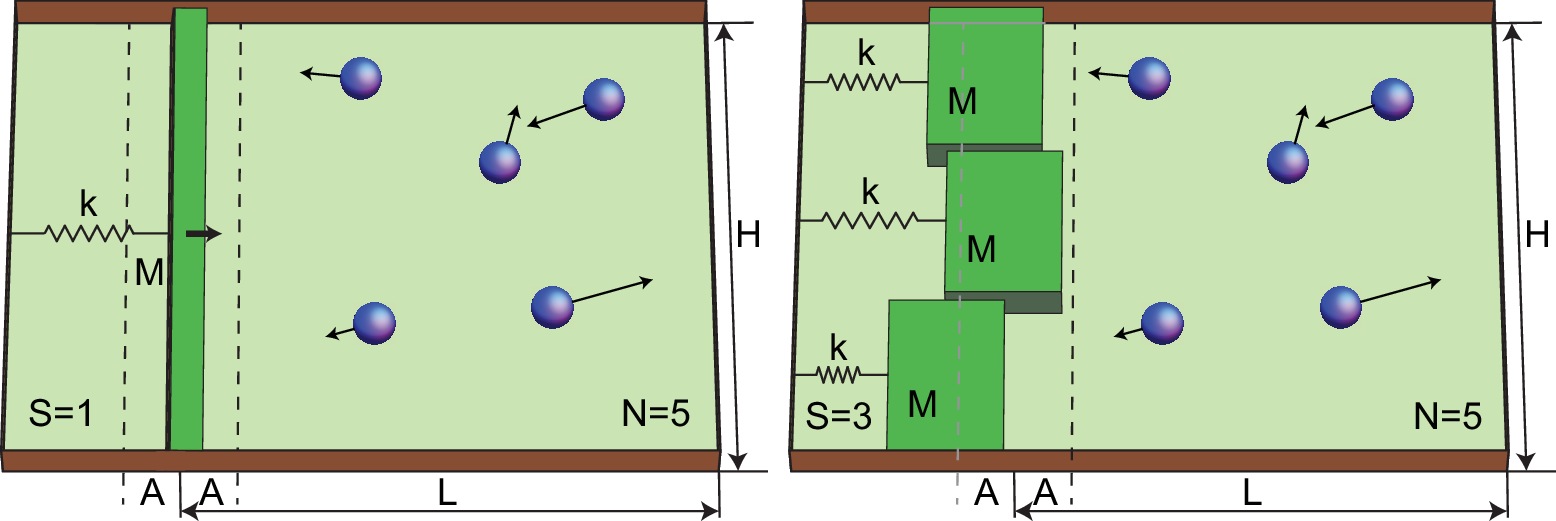}\\
 \caption{A finite number $N$ of colliding particles in the rectangular container of length $L$ and height $H$. One side of the container consists of $S$ mobile elements ($S=1$ and $S=3$ are shown). All the mobile elements are of mass $M$ and holden by springs of stiffness $k$. Between collisions with particles they oscillate with a frequency $w=\sqrt{\frac{k}{M}}$ and some current amplitude $A_i(t) \leq A$.}
 \label{eqstat-fig1}
\end{figure}

We will consider a two-dimensional motion of a finite number $N$ of colliding particles in a finite-sized rectangular container. The particles will be of a round shape, with radii $r_i$ and masses $m_i$. Their motion between collisions will be uniform and rectilinear, all collisions absolutely elastic. The container will be of length $L$ and height $H$. One of its walls will consist of $S$ equivalent movable parts of mass $M$, holden by springs of stiffness $k$. The general view of this system is shown in Fig.\ref{eqstat-fig1}. Between collisions with particles, each mobile part oscillates freely with a frequency $w=\sqrt{\frac{k}{M}}$. After each collision, the energy, amplitude and phase of the oscillation change. Since the total system energy remains finite and equal to a constant value $E_{tot}$, the coordinates $X_i$ of the mobile elements always stay within the range $|X_i| \leq A=\sqrt{\frac{2 E_{tot}}{k}}$.

The complete system's phase space will be the $4 N+2 S$-dimensional space of variables $(P_1,..,P_S$, $p_{x 1},..,p_{x N}$, $p_{y 1},..,p_{y N}$ , $X_1,..,X_S$, $x_1,..,x_N$, $y_1,..,y_N)$, where $(p_{x i},p_{y i})$, $(x_i,y_i)$, $P_i$, $X_i$ are the momenta and coordinates of the $i$th particle and surface part respectively. The current state of the system is represented by some point in this phase space. Since the total energy of the system is conserved, the representing point during evolution is always located on the surface of constant energy $E=E_{tot}$. As is known, the density of filling of this surface with trajectory is uniform for a gas of identical particles. For a gas of particles of different masses, the filling density is also uniform with respect to the special measure, which is called gradient or ergodic. According to this measure, hypervolume of an elementary hypersurface part is determined as $d \Omega = \frac{d \Sigma}{|grad~E|}$, where $d \Sigma$ is the hypervolume of this elementary part according to the usual Euclidean measure. The derivation of the gradient measure from Liouville's theorem, as well as the derivation of this theorem from the Hamiltons equations of motion, can be found, for example, in \cite{eqstat-Hin}. Thus, the density of filling of the surface of constant energy is:

\begin{equation}\label{eqstat-eq3}
\rho = \frac{\mathit{const}}{|grad~E|}
\end{equation}

This is a probability density per unit hypervolume of a curvilinear $4 N+2 S -1$-dimensional hypersurface. We will write the equation of this surface as:

\begin{equation}\label{eqstat-eq4}
P_1 = \pm \sqrt{M} \sqrt{2 E_{tot}-\underset{i=1}{\stackrel{N}\sum} \frac{(p_{x i}^2+p_{y i}^2)}{m_i}-k X_1^2-\underset{j=2}{\stackrel{S}\sum} (\frac{P^2_j}{M}+k X_j^2)}
\end{equation}

\begin{figure}
 \centering
 \includegraphics[width=6 cm]{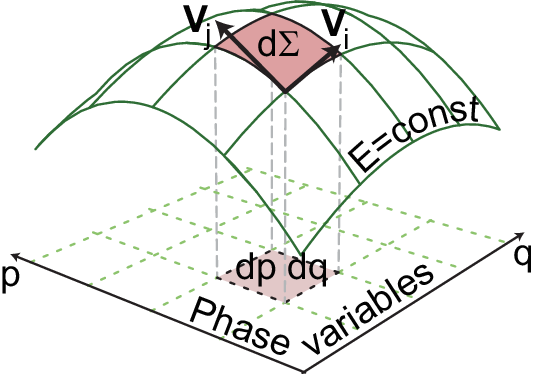}\\
 \caption{The elementary volume $d \Sigma=|\mathbf{V}| dP_2..dy_N$, spanned over a set of vectors $\mathbf{v_i}$, and its projection $dP_2..dy_N$ onto the space of independent phase variables.}
 \label{eqstat-fig6}
\end{figure}

Thus, we choose $P_1$ as a dependent variable, which is determined (up to the sign) by all the other independent phase variables. Further, we are going to find the relationship between the elementary volume $d \Sigma$ of the energy hypersurface and the volume of its projection onto the space of independent phase variables (see Fig.\ref{eqstat-fig6}). Each vector $(0,...,0,dp_i,0,...,0)$ is a projection of the vector $\mathbf{v_i} dp_i=(\frac{dP_1} {dp_i}dp_i,0,...,0,dp_i,0,..,0)$ that lies on the energy hypersurface. Here $p_i$ is any of the independent phase variables, standing in its place. The elementary part of the space of independent phase variables $dP_2..dy_N$ is the orthogonal projection of the multidimensional parallelepiped $d \Sigma$, spanned over the vectors $\mathbf{v_i} dp_i$. The volume of this parallelepiped is equal to the modulus of the vector $\mathbf{V}=\mathbf{v_1} \times ... \times \mathbf{v_{4N+2S-1}}$, which is:

\begin{equation}\label{eqstat-eq6}
\mathbf{V}=(-1,-\frac{P_2}{P_1},..,-\frac{P_S}{P_1},-\frac{M p_{x 1}}{m_1 P_1},..,-\frac{M p_{y N}}{m_N P_1},-\frac{M k X_1}{P_1},..,-\frac{M k X_S}{P_1},0,..,0) = - \frac{M}{P_1} grad~E
\end{equation}

Then, for the filling density of the projection of energy hypersurface we have:

\begin{equation}\label{eqstat-eq7}
\rho_{proj}=const \frac{|\mathbf{V}|}{|grad \; E|}=\frac{const}{|P_1|}=\frac{const}{\sqrt{2 E_{tot}-k X_1^2-\underset{i=1}{\stackrel{N}\sum} \frac{(p_{x i}^2+p_{y i}^2)}{m_i}-\underset{j=2}{\stackrel{S}\sum} (\frac{P^2_j}{M}+k X_j^2)}}
\end{equation}

This is the probability density that the coordinates of gas particles will be $x_1..x_N,y_1,..,y_N$, the surface elements coordinates will be $X_1..X_S$, the particle momenta components will be $p_{x 1}..p_{x N} ,p_{y 1}..p_{y N}$ and the surface elements momenta will be $P_2..P_S$. Further we will integrate the probability density Eq.\ref{eqstat-eq7} over different sets of phase variables to obtain the statistical distributions of interest. The regions of integration will be finite, since the law of energy conservation imposes restrictions on the maximum possible particle momenta, mobile elements momenta and displacements. Particle coordinates are limited by the container walls. Thus, all the phase variables change within some finite limits. We will use one essential assumption that a typical system's trajectory fills entirely the region within these limits. This is a widely used assumption, when a gas of colliding particles is considered. We confirmed its validity by comparison of theoretical and simulation results.

\section{Coordinates and momenta distributions}

Let us first consider in detail the most simple case of only one mobile surface element $S=1$ or a mobile wall Fig.\ref{eqstat-fig1}(a). In order to obtain the distribution of the wall coordinates $p(X)$, the expression Eq.\ref{eqstat-eq7} must be integrated over all the particle coordinates and momenta:

\begin{equation}\label{eqstat-eq8}
p(X)=\int_{p_i=0}^{p_{i max}} \int_{x_i=X}^{L} \frac{const \underset{i=1} {\stackrel{N}\prod} p_i }{\sqrt{2 E_{tot}-k X^2-\underset{i=1}{\stackrel{N}\sum} \frac{p_i^2}{m_i}}}  dp_1..dp_N dx_1..dx_N
\end{equation}

where:

\begin{equation}\label{eqstat-eq9}
\begin{array}{l}
p_{1\,max}=\sqrt{m_1}\sqrt{2 E_{tot}-k X^2}\\
p_{i\,max}=\sqrt{m_i}\sqrt{2 E_{tot}-k X^2-\underset{j=1}{\stackrel{i-1}\sum}\frac{p_j^2}{m_j}}, \qquad i=2..N\\
\end{array}
\end{equation}

Here a transition is made from the momenta components to the momenta modulus and the angles of the particles motion. The result of integration over the angles and the $y$-coordinates of particles is a part of the constant. The integration over momenta can be done with the help of the following mathematical relation:

\begin{equation}\label{eqstat-eq10}
\int_{p_j=0}^{\sqrt{m_j}\sqrt{2 E_{tot}-k X^2-\underset{i=1}{\stackrel{j-1}\sum}\frac{p_i^2}{m_i}}} (2 E_{tot}-k X^2-\underset{i=1}{\stackrel{j}\sum} \frac{p_i^2}{m_i})^{N-j-1/2} p_j \: dp_j = (2 E_{tot}-k X^2-\underset{i=1}{\stackrel{j-1}\sum} \frac{p_i^2}{m_i})^{N-j+1/2}
\end{equation}

Each result of the momenta integration is the integrand with removed variable of integration and increased degree. Each integration over $x$-coordinates adds the factor $\int_{x_j=X}^{L} dx_j = L-X$. Let us underline here that due to the inequalities $x_i \geq X$, the region of particle coordinates is not the entire $x_i \in [-A,L]$, but $x_i \in [X,L]$.

\begin{figure}
 \centering
 \includegraphics[width=7 cm]{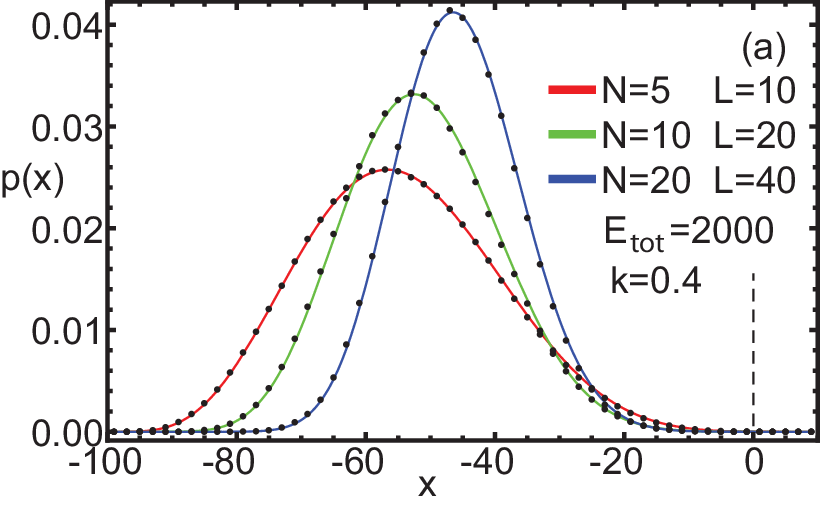}
 \includegraphics[width=7 cm]{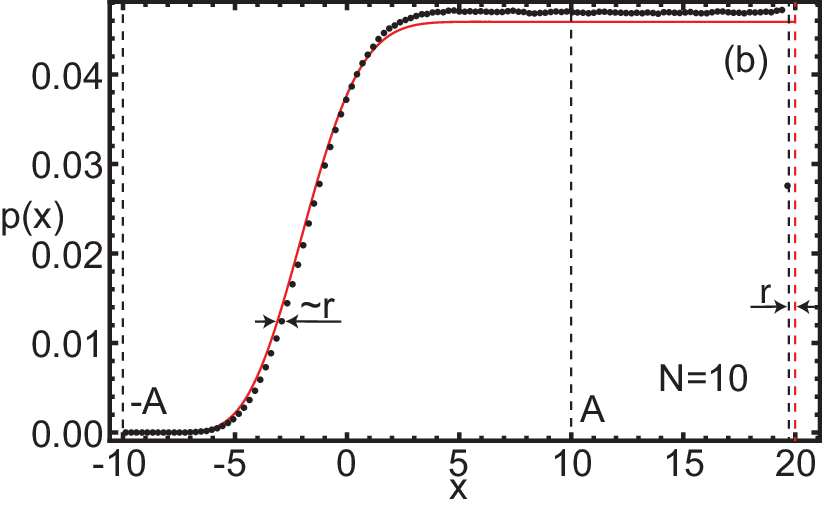}\\
  \includegraphics[width=7 cm]{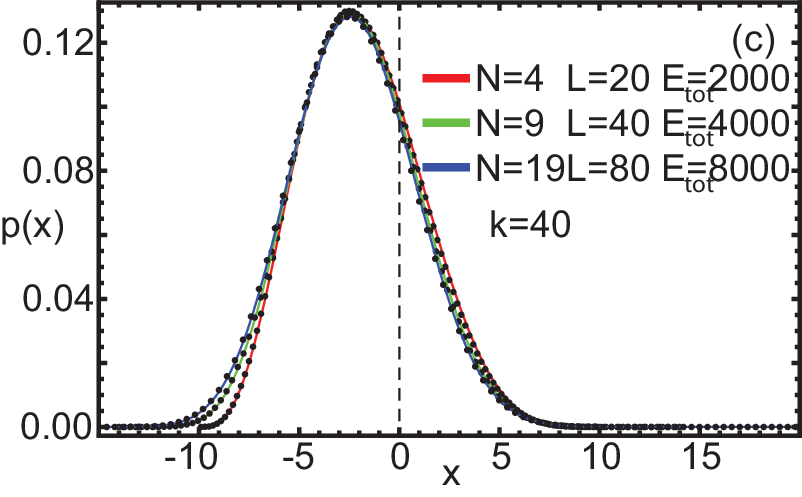}
 \includegraphics[width=7 cm]{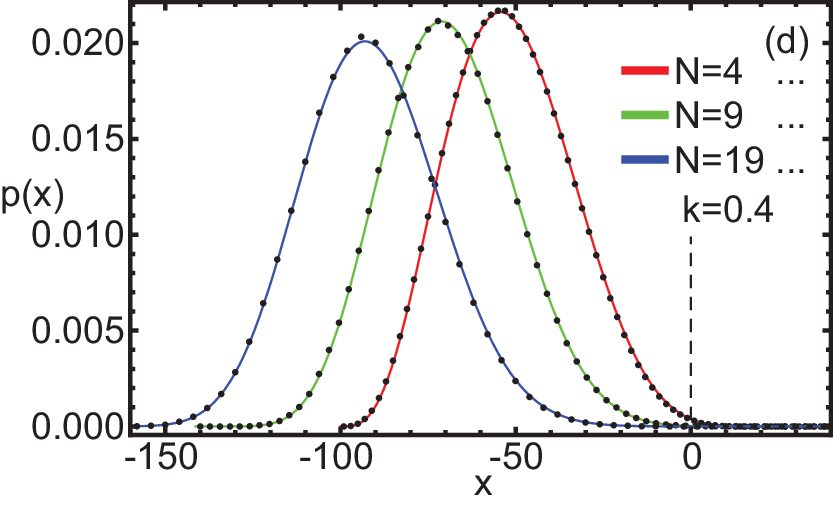}\\
 \caption{Distributions of coordinates of the mobile wall and the gas particles. Continuous curves correspond to theoretical distributions Eq.\ref{eqstat-eq12} and Eq.\ref{eqstat-eq23}, dots show the results of numerical simulation of particle motion. In all cases, particle masses $m_i=1$, particle sizes $r_i=0.3$, wall mass $M=10$. (a) Typical distributions of the wall coordinate. (b) Typical distribution of gas particle coordinates. The experimental distribution slightly differs from the theoretical one due to the finiteness of the particles sizes. (c) If the effective concentration and average energy of gas particles are kept almost equal, the distribution of wall coordinates practically does not depend on the number of particles. (d) The previous case with the only difference in the reduced spring stiffness. The mean displacement of the wall significantly changes the effective volume and, hence, the concentration of particles.}
 \label{eqstat-fig2}
\end{figure}

As a result, we obtain the following distribution of a mobile wall coordinates:

\begin{equation}\label{eqstat-eq12}
p(X)= C_1 (L-X)^N (2 E_{tot}-k X^2)^{N-1/2} = C (1-\frac{X}{L})^N (1-\frac{X^2}{A^2})^{N-1/2}
\end{equation}

where $C$ is a normalization constant equal to:

\begin{equation}\label{eqstat-eq13}
C=
\begin{cases}
\frac{1}{A \sqrt{\pi}} \; \frac{\Gamma(N+1)}{\Gamma(N+\frac{1}{2})} \; \frac{1}{_2F_1(\{\frac{1-N}{2},-\frac{N}{2}\},N+1,\frac{A^2}{L^2})} \qquad \qquad \qquad \qquad \qquad \qquad  A \leq L\\
\frac{1}{A \sqrt{\pi}} \; \frac{\Gamma(2N+\frac{3}{2})}{(2N)!} \; \frac{(4 A^2 L)^N }{  (A-L)^{N-\frac{1}{2}} (A+L)^{2N+\frac{1}{2}}} \; \frac{1}{_2F_1(\{\frac{1}{2}-N,N+1\},2N+\frac{3}{2},\frac{L+A}{L-A})} \qquad A>L \\
\end{cases}
\end{equation}

The $_pF_q(\textbf{a};\textbf{b};z)$ here is a generalized hypergeometric function, defined as:

\begin{equation}
_pF_q(\textbf{a};\textbf{b};z) = \underset{k=0} {\stackrel{\infty}\sum} \frac{(a_1)_k ... (a_p)_k}{(b_1)_k ... (b_q)_k} \frac{z^k}{k!}; \quad (a)_n =\underset{i=0} {\stackrel{n-1}\prod} (a+i) .
\end{equation}

A comparison of the distribution Eq.\ref{eqstat-eq12} with the results of numerical modeling is shown at Fig.\ref{eqstat-fig2}(a,c,d). The figure Fig.\ref{eqstat-fig2}(a) shows distributions for different combinations of $N$ and $L$ at the same ratio $\frac{N}{L}$ and total energy $E_{tot}$. The average location of the wall is far from the equilibrium position $X=0$ due to the spring of small stiffness being compressed under the gas pressure. The more particles are inside the vessel, the less is the average spring compression at constant $E_{tot}$ and $\frac{N}{L}$. This is a consequence of the decrease in the mean particle energy. The distribution will remain practically unchanged with constant ratios $\frac{E_{tot}}{N+1}$ and $\frac{N+1}{L}$, as shown in Fig.\ref{eqstat-fig2}(c). The wall should be accounted as an additional particle and the spring must be sufficiently rigid. If to choose a non-stiff spring with other parameters being the same, as shown in Fig.\ref{eqstat-fig2}(d), then the concentration of particles will not remain the same and distributions will differ significantly.

Further we will consider only the case of stiff enough container walls $A=\sqrt{\frac{2 E_{tot}}{k}} \leq L$.

Let us now calculate the distribution of $x$-coordinates of gas particles. The procedure is similar to the previous one, but the integration over coordinate of one of particles should be changed to the integration over the wall coordinate $X$. The excluded particle coordinate, $x_1$ for example, appears within the limits of integration over $X$, but only at $x_1<A$, when $-A \leq X \leq x_1$. In the case of $x_1>A$, the limits of integration over $X$ are $-A \leq X \leq A$. As a result, at $x_1>A$ the particle coordinate $x_1$ does not appear anywhere and the integration result is a constant. Likewise, the distribution of the $y$-coordinate is constant. Thus, the probability density $p(x_1)$ consists of two sites. The first site is:

\begin{equation}\label{eqstat-eq23-0}
\begin{array}{l}
p(x_1)=\underset{X=-A}{\stackrel{x_1}\int} \int_{p_i=0}^{p_{i max}} \int_{x_i=X}^{L} \frac{const \underset{i=1} {\stackrel{N}\prod} p_i }{\sqrt{2 E_{tot}-k X^2-\underset{i=1}{\stackrel{N}\sum} \frac{p_i^2}{m_i}}}  dX dp_1..dp_N dx_2..dx_N = \\
=const \underset{X=-A}{\stackrel{x_1}\int} (1-\frac{X}{L})^{N-1} (1-\frac{X^2}{A^2})^{N-\frac{1}{2}} dX \qquad \qquad x_1 \in (-A,A)\\
\end{array}
\end{equation}

which gives:

\begin{equation}\label{eqstat-eq23-1}
\begin{array}{l}
p(x_1)=const (\frac{\sqrt{\pi} (2N)! (A+L)^N}{\Gamma (2N+\frac{1}{2})}\,_2F_1 (\{\frac{1}{2}-N,N\},2N+\frac{1}{2},\frac{L+A}{L-A})-\qquad\qquad\qquad\\
\qquad\qquad\qquad-2^{2N} (L-x_1)^N F_1 (N,\{ \frac{1}{2} - N, \frac{1}{2} - N\}, N + 1,\frac{L-x_1}{L+A},\frac{L-x_1}{L-A})) \qquad x_1 \in (-A,A)\\
\end{array}
\end{equation}
Here $F_1 (a,\{b_1,b_2\},c,x,y)$ is a hypergeometric Appel function. It is difficult to work with even numerically. In practice, it is easier to numerically calculate the preceding definite integral. The second distribution site is a constant:

\[C_x=\underset{X=-A}{\stackrel{A}\int} \!\!\! (1-\frac{X}{L})^{N-1} (1-\frac{X^2}{A^2})^{N-\frac{1}{2}} dX=\frac{A \sqrt{\pi} \, \Gamma(N+\frac{1}{2}) \; _2F_1(\{\frac{1-N}{2},1-\frac{N}{2}\},N+1,\frac{A^2}{L^2})}{\Gamma(N+1)} \]

In the result, for the distribution of $x$-coordinates of particles we write:

\begin{equation}\label{eqstat-eq23}
p(x_1)=const
\begin{cases}
\underset{X=-A}{\stackrel{x_1}\int} \!\!\! (1-\frac{X}{L})^{N-1} (1-\frac{X^2}{A^2})^{N-\frac{1}{2}} dX \qquad \,\,\, x_1 \in (-A,A)\\
\,\, C_x \quad \quad \quad \quad \quad \quad \quad \quad \quad \quad \quad \quad \quad \quad \quad x_1 \in (A,L) \\
\end{cases}
\end{equation}

A comparison of the distribution Eq.\ref{eqstat-eq23} with the results of numerical simulation is shown in Fig.\ref{eqstat-fig2}(b). The slight difference between the experimental and theoretical distributions is due to the finiteness of particle sizes. In particular, the experimental distribution ends on the value $L-r$, while the theoretical one proceeds to $L$.

Let us now consider the energy and momentum distributions of gas particles and the wall. These distributions proceed to some finite values, and then they are exactly zero. Integrating the probability density Eq.\ref{eqstat-eq7}, for the momentum distribution $p(p_1)$ of the first particle (at $A \leq L$) we have obtained:

\begin{equation}\label{eqstat-eq143}
\begin{array}{l}
p(p_1)=\int_{X=-\sqrt{A^2-\frac{p_1^2}{k m_1}}}^{\sqrt{A^2-\frac{p_1^2}{k m_1}}} \int_{p_i=0}^{p_{i max}} \int_{x_i=X}^{L} \frac{const \underset{i=1} {\stackrel{N}\prod} p_i }{\sqrt{2 E_{tot}-k X^2-\underset{i=1}{\stackrel{N}\sum} \frac{p_i^2}{m_i}}}  dX dp_2..dp_N dx_1..dx_N = \\
= const \; p_1 \: \int_{X=-\sqrt{A^2-\frac{p_1^2}{k m_1}}}^{\sqrt{A^2-\frac{p_1^2}{k m_1}}} (L-X)^N (2 E_{tot}-k X^2-\frac{p_1^2}{m_1})^{N-3/2} dX \end{array}
\end{equation}

from where:

\begin{equation}\label{eqstat-eq144}
\begin{array}{l}
p(p_1)=C \; p_1 \:(E_{tot}-\frac{p_1^2}{2 m_1})^{N-1} \: _2F_1 (\frac{1-N}{2},-\frac{N}{2},N,\frac{2 E_{tot} m_1 - p_1^2}{m_1 k L^2});\\
\\
\quad \:\: C=\frac{N}{E_{tot}^N \: _2F_1 (\{\frac{1-N}{2},-\frac{N}{2}\},N+1,\frac{2 E_{tot}}{k L^2})}
\end{array}
\end{equation}

The integration limits here and the integration procedure are the same as previously (see Eq.\ref{eqstat-eq9}, Eq.\ref{eqstat-eq10}). From the momentum distribution, it is easy to obtain the distribution of kinetic energy of gas particles:

\begin{equation}\label{eqstat-eq15}
\begin{array}{l}
p(E_{P K})=C (E_{tot}-E_{P K})^{N-1} \: _2F_1 (\{\frac{1-N}{2},-\frac{N}{2}\},N,2 \frac{E_{tot}-E_{P K}}{k L^2});\\
\\
\qquad \:\: C=\frac{N}{E_{tot}^N \: _2F_1 (\{\frac{1-N}{2},-\frac{N}{2}\},N+1,\frac{2 E_{tot}}{k L^2})}
\end{array}
\end{equation}

Since this distribution does not depend on particles masses, all particles share the same energy distribution. But, in contrast to the Boltzmann distribution, it includes the dependence on the combination $k L^2$. In this way, the distribution of energy of gas particles appears to depend on the geometry of the container walls and their stiffness.

Similarly, for the distribution of the kinetic energy of the wall $E_{W K} =E_{tot}-\frac{k X^2}{2}-\underset{i=1}{\stackrel{N}\sum} \frac{ (p_{x i}^2+p_{y i}^2)}{2 m_i}$ we have obtained:

\begin{equation}\label{eqstat-eq22}
\begin{array}{l}
p(E_{W K}) = C \frac{(E_{tot}-E_{W K})^{N-\frac{1}{2}}}{\sqrt{E_{W K}}} \: _2F_1 (\{-\frac{N}{2},\frac{1-N}{2}\},N+\frac{1}{2},2 \frac{E_{tot}-E_{W K}}{k L^2})\\
\\
\qquad \;\;\, C=\frac{\Gamma(N+1)}{\sqrt{\pi} \Gamma(N+\frac{1}{2}) E_{tot}^N \: _2F_1 (\{\frac{1-N}{2},-\frac{N}{2}\},N+1,\frac{2 E_{tot}}{k L^2})}
\end{array}
\end{equation}

\begin{figure}
 \centering
 \includegraphics[width=7 cm]{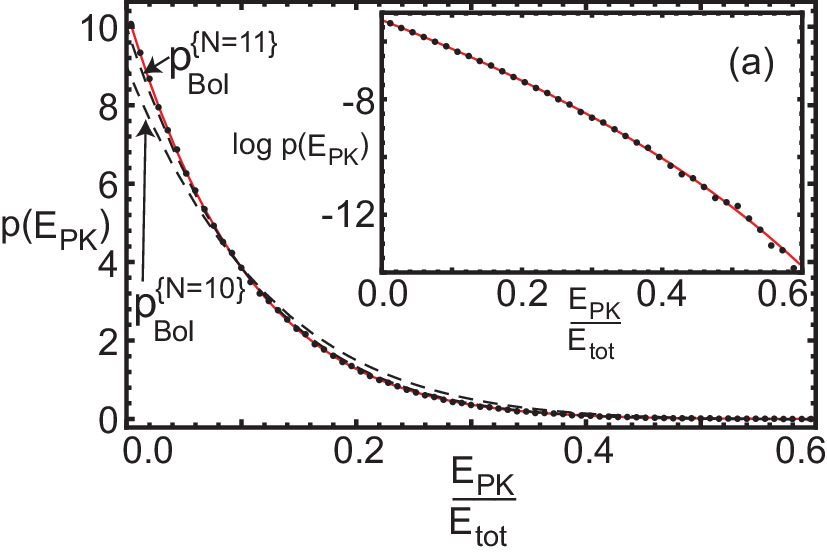}
 \includegraphics[width=7 cm]{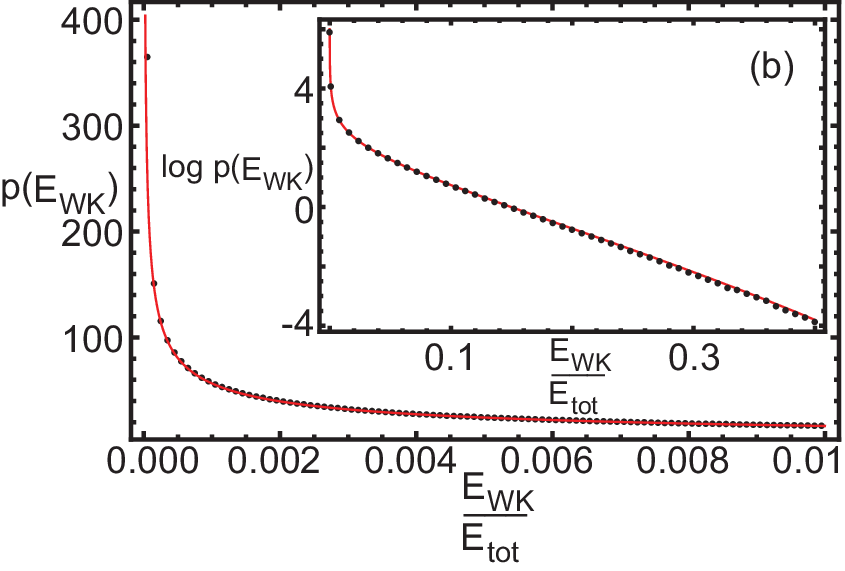}\\
 \caption{Kinetic energy distributions of (a) the gas particles, (b) the spring wall, in case of $N=10$ particles in the vessel. Dots show the numerical modelling results, continuous curves show distributions Eq.\ref{eqstat-eq15} and Eq.\ref{eqstat-eq22}. The inserts show the same distributions in a logarithmic scale. (a) For comparison, Boltzmann distributions for $N=10$ and $N=11$ particles are shown. (b) Distribution of the wall's kinetic energy is concentrated at zero energies region.}
 \label{eqstat-fig3}
\end{figure}

A comparison of the distributions Eq.\ref{eqstat-eq15} and Eq.\ref{eqstat-eq22} with the results of numerical modeling is shown in Fig.\ref{eqstat-fig3}. The obtained energy distribution for $N$ particles appears to be close to the Boltzmann distribution for $N+1$ particles, as shown in Fig.\ref{eqstat-fig3}(a). The distribution of the kinetic energy of the mobile wall is very different. The main part of this distribution is concentrated at zero energies region, i.e. the kinetic energy of the wall is usually close to zero. However, the tail of this distribution is power-like and decays slowly, as shown in the insert in Fig.\ref{eqstat-fig3}(b). Below we will show, that the average kinetic energies of the gas and the surface parts per degree of freedom are exactly equal.

Thus, the main statistical distributions for the modeled system were explicitly obtained. The particle $x$-coordinates distribution is obtained up to the normalization constant. All the distributions are independent of any masses. All internal particles share the same energy distribution, which is geometry-dependent. While this distribution is close to the usual Boltzmann's distribution, the kinetic energy distribution of the surface element is new and far from Boltzmann's one.

\section{Distribution of energy over the degrees of freedom}

Let us now consider the mean energies of the internal particles and the mobile wall. The average potential energy $\overline{\mathcal{E}}_P$ of the wall can be calculated from the distribution of coordinates $p(X)$ (Eq.\ref{eqstat-eq12}):

\begin{equation}\label{eqstat-eq14}
\overline{\mathcal{E}}_P=\underset{X=-A}{\stackrel{A}\int} \frac{k X^2}{2} p(X) dX = \frac{E_{tot}}{2(N+1)} \frac{_3F_2(\{\frac{3}{2},\frac{1-N}{2},-\frac{N}{2}\},\{\frac{1}{2},N+2\},\frac{A^2}{L^2})}{_2F_1(\{\frac{1-N}{2},-\frac{N}{2}\},N+1,\frac{A^2}{L^2})}
\end{equation}

The first fraction in this expression corresponds to the energy equipartition level. Second fraction is always greater than unity, tending to it with $\frac{A}{L} \rightarrow 0$. Thus, the mean potential energy of the wall is always above the equipartition level, which means the distribution of energy over the degrees of freedom is not uniform.

The wall's average potential energy appears to depend on the geometry of the vessel. Why this should be the case can be understood from simple qualitative considerations. The pressure of the gas is determined not only by the mean energy of the gas particles, but also by their concentration, which depends on the vessel length $L$. The wall's mean displacement is directly proportional to the pressure. Therefore, the smaller $L$ will be (at constant kinetic energy level), the greater will be the concentration, pressure and wall's mean displacement. The growth of the average displacement leads to the growth of the average potential energy $\overline{\mathcal{E}}_P$. For this reason, $\overline{\mathcal{E}}_P$ is dependent on the vessel length $L$ (but not the height). It is interesting to note that the above qualitative considerations remain valid even if the system is attached to a thermostat. The average wall's displacement and, hence, its average potential energy will still depend on the gas pressure, not the temperature only. Such system is a counterexample to the version of the equipartition theorem given in Kubo's book \cite{eqstat-Kub}.

This new counterexample shows that the conditions of applicability of the equipartition theorem still need to be clarified. It is well known that it does not apply to all Hamiltonian systems. Some of the counterexamples, such as a well-known chain of coupled oscillators or an ideal gas in a round vessel \cite{eqstat-my,eqstat-my1} are based on the existence of additional integrals of motion. If a system has any other conserved quantities except for its energy, the trajectory will not fill the surface of constant energy entirely. The system considered above belongs to a different kind of counterexamples, since it does not have additional integrals of motion. But its phase variables are bound with inequalities, which also limits the filled phase space area. And the proofs of the equipartition theorem are based on the integration over the entire energy surface. Its fulfillment is one of the theorem's requirements, which is usually not stated explicitly. For many Hamiltonian systems this condition is not satisfied.

The lack of equipartition in the considered system is a consequence of relations $x_i>X$. The particles should be located inside the vessel. For this reason, during derivation of the distribution of wall coordinates, the integration region was not the entire range of coordinates $-A \leq x_i \leq L$. One can integrate over this range, as is assumed in the proofs of the energy equipartition theorem. Then, an incorrect distribution $p(X) = C (1-\frac{X^2} {A^2})^{N-1/2}$ will be obtained. It corresponds to the energy equipartition. However, the regions $-A<x_i<X$ should be excluded since they correspond to particles being outside the vessel. In the result, the energy equipartition appears to be violated with the mean potential energy of the surface being higher than the equipartition level. This opens a possibility to theoretically introduce a component of the surface energy, associated with the internal pressure the surface holds in.

Let us now calculate the average kinetic energies of the particles and the wall:

\begin{equation}\label{eqstat-eq16}
\begin{array}{l}
\overline{\mathcal{E}}_{P K}=\underset{E_{P K}=0}{\stackrel{E_{tot}}\int} E_{P K} \: p(E_{P K}) dE_{P K}\\
\overline{\mathcal{E}}_{W K}=\underset{E_{W K}=0}{\stackrel{E_{tot}}\int} E_{W K} \: p(E_{W K}) dE_{W K}
\end{array}
\end{equation}

Where $p(E_{P K})$ and $p(E_{W K})$ are given by Eq.\ref{eqstat-eq15} and Eq.\ref{eqstat-eq22}. After calculation of the integrals, we have obtained:

\begin{equation}\label{eqstat-eq18}
\overline{\mathcal{E}}_{W K}=\frac{\overline{\mathcal{E}}_{P K}}{2}=\frac{E_{tot}}{2 N+1}\left(1-\frac{1}{2(N+1)} \frac{_3F_2(\{\frac{3}{2},\frac{1-N}{2},-\frac{N}{2}\},\{\frac{1}{2},N+2\},\frac{A^2}{L^2})}{_2F_1(\{\frac{1-N}{2},-\frac{N}{2}\},N+1,\frac{A^2}{L^2})}\right)
\end{equation}

\begin{figure}
 \centering
 \includegraphics[width=7 cm]{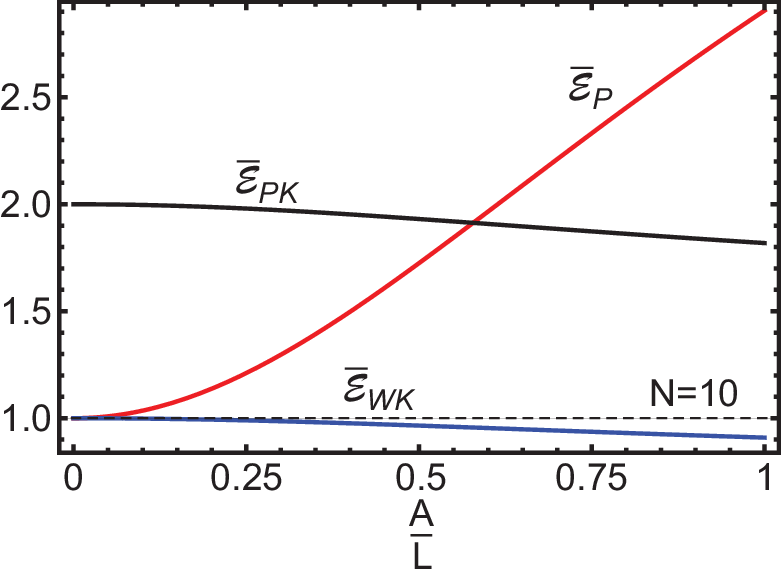}\\
 \caption{Distribution of energy over the system's degrees of freedom. The mean kinetic energy of a gas particle $\overline{\mathcal{E}}_{P K}$, mean kinetic $\overline{\mathcal{E}}_{W K}$ and potential $\overline{\mathcal{E}}_P$ energy of the wall are shown in dependence on $\frac{A}{L}$. The dotted line shows the energy equipartition level. The mean kinetic energy of a particle per degree of freedom $\frac{\overline{\mathcal{E}}_{P K}}{2}$ coincides with $\overline{\mathcal{E}}_{W K}$. The smaller the length of the vessel $L$, the greater the average potential energy of the wall $\overline{\mathcal{E}}_P$.}
 \label{eqstat-fig5}
\end{figure}

Thus, the kinetic energy appears to be equally partitioned. Despite the very different energy distributions, mean energies of particles and mobile wall are equal, per degree of freedom. The dependencies of the average energies $\overline{\mathcal{E}}_{W K}$, $\overline{\mathcal{E}}_P$ and $\overline{\mathcal{E}}_{P K}$ on the parameter $\frac{A}{L}$ are shown in Fig.\ref{eqstat-fig5}. The average kinetic energy is slightly lower than the energy equipartition level. The average potential energy of the wall can be significantly higher than the equipartition level. With the limit passage $k \rightarrow \infty$, the overall energy distribution becomes uniform.

\section{The equations of state}

Let us now consider the equation of state of an ideal gas of a finite number of particles. We will obtain it as the dependence of the gas pressure $P_{gas}$ on the parameters, from the condition that the spring holds a mobile wall, compensating for the gas pressure. There are two forces acting on the mobile wall of a container: the force from the compressed spring and the force that occurs during particles reflections. The average value of the latter per unit area is the gas pressure. The long time average of the total force acting on the wall must be equal to zero $P_{gas} H + k \overline{X} = 0$. The mean displacement $\overline{X}$ can be easily calculated from the $p(X)$ distribution (Eq.\ref{eqstat-eq12}). In this simple way, the gas pressure may be strictly obtained, without direct calculations of the momentum transfer in collisions, distributions of collision times, their correlations, etc.

\begin{equation}\label{eqstat-eq19}
P_{gas} H = - k \overline{X} = - k \underset{x=-A}{\stackrel{A}\int} X p(X) dX = \frac{E_{tot} N}{L (N+1)} \frac{ _2F_1(\{\frac{1-N}{2},1-\frac{N}{2}\},N+2,\frac{A^2}{L^2})}{_2F_1(\{\frac{1-N}{2},-\frac{N}{2}\},N+1,\frac{A^2}{L^2})}
\end{equation}

\begin{figure}
 \centering
 \includegraphics[width=7 cm]{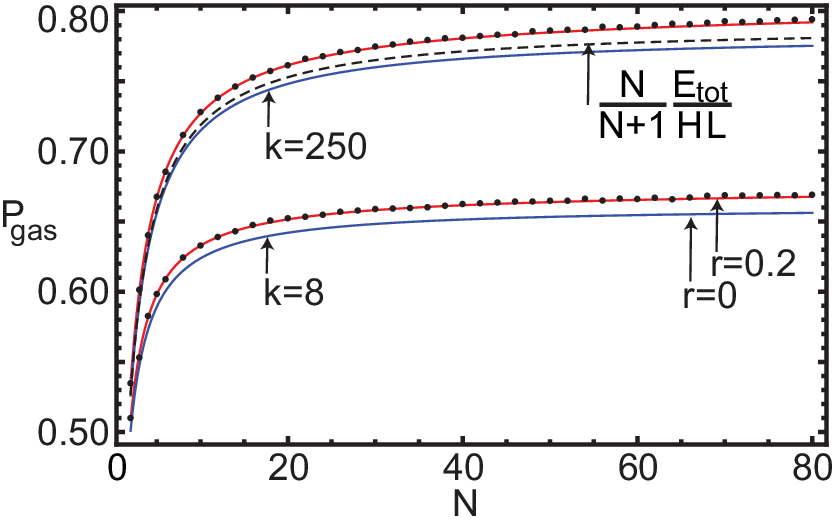}\\
 \caption{The gas pressure in dependence on the number of particles in a vessel with a spring wall. Continuous curves show the theoretical dependencies for sizeless (Eq.\ref{eqstat-eq19}) and finite-sized (Eq.\ref{eqstat-eq19-1}) particles. The dots show the results of numerical simulation. With the spring stiffness tending to infinity, the gas pressure tends to $P=\frac{N}{N+1} \frac{E_{tot}}{H\,L}$ (shown with dotted line). System parameters are: $E_{tot}=2000$, $L=40$, $H=20 \sqrt{10}$, $m_i=1$, $M=10$, $r_i=0.2$, $k=8$ or $k=250$.}
 \label{eqstat-fig4}
\end{figure}

Thus, the exact equation of state of a finite system with one mobile surface element can be obtained explicitly. It depends non-trivially on the wall's stiffness and other parameters. Comparison of the theoretical pressure value Eq.\ref{eqstat-eq19} with the results of numerical modeling is shown in Fig.\ref{eqstat-fig4}. The pressure was determined numerically as the total momentum, transferred to the wall in collisions with particles, divided by the evolution time and the wall size. It was determined for the mobile wall and the stationary vertical and horizontal walls (its part from $A$ to $L$). All three pressures coincided within the accuracy of numerical calculations. But the experimental pressure value turned out to be noticeably different from Eq.\ref{eqstat-eq19}. This difference is insignificant for a small number of particles, but grows with the increase in their number, as shown in Fig.\ref{eqstat-fig4}(b). This is found to be a consequence of the dependence of pressure on the size of gas particles. The particle size can not be set too small in modelling, because collisions have to equalise horizontal and vertical pressures. But it can be accounted for theoretically. For that it is only necessary to change the range of possible particle coordinates during integration of Eq.\ref{eqstat-eq7}. It is easy to see that in the result of new integration, the factor $(H(L-X))^N$ in the distribution of wall coordinates will change to $((H-2r)(L-X-2r)-4 N r^2)^N$, where $4 N r^2$ is the approximate value of the excluded volume. After calculation of a new average wall position, for the pressure of the gas of finite-sized particles of radius $r$ we obtain:

\begin{equation}\label{eqstat-eq19-1}
P^r_{gas} \approx \frac{E_{tot} N}{(N+1)} \frac{1}{(H-2r)(L-2r)-4 N r^2} \frac{ _2F_1(\{\frac{1-N}{2},1-\frac{N}{2}\},N+2,\frac{A^2 (H-2r)^2 }{((H-2r)(L-2r)-4 N r^2)^2})}{_2F_1(\{\frac{1-N}{2},-\frac{N}{2}\},N+1,\frac{A^2 (H-2r)^2}{((H-2r)(L-2r)-4 N r^2)^2})}
\end{equation}

This expression is valid at $H L>>4 N r^2$. It well agrees with the simulation results (see Fig.\ref{eqstat-fig4}(b)). Thus, the numerical modelling confirms the analytical expression Eq.\ref{eqstat-eq19-1}. The account for the particle sizes was necessary only to assure that. Since an ideal gas is a gas of negligible size particles, we will proceed further with $r=0$.

Let us now calculate the equation of state for an ideal gas in an ordinary stationary container. To transfer to the container with stationary walls, the spring stiffness should be set to infinity $k \rightarrow \infty$ ($A \rightarrow 0$). The increase in the mass of the mobile wall does not affect its coordinate or energy distributions, but only the time of their establishment. It is interesting to note that with $k \rightarrow \infty$, the wall's momentum remains non-zero, since the mean kinetic energy of the wall $\frac{\overline{P^2}}{2 M}$ tends to the equipartition level, not to zero. This is possible because of the tending to infinity frequency $w=\sqrt{\frac{k}{M}}$ of switching of directions of wall's motion. The gas pressure in the stationary walls limit will be:

\begin{equation}\label{eqstat-eq21}
P_{gas} \; \underset{k \rightarrow \infty}{\rightarrow} \; \frac{N}{N+1}\frac{E_{tot}}{H L} = 2 \frac{N}{V} \frac{E_{tot}}{2(N+1)} = 2 n \overline{\mathcal{E}}_K \\
\end{equation}

Thus, the equation of state for a gas of a finite number of particles appears to differ from the usual ideal gas one only in the change of $\frac{1}{2}kT$ to $\overline{\mathcal{E}}_K$, if the container walls are perfectly stiff and there are no influence of the surface.

In the thermodynamic limit $N \rightarrow \infty$ at $\frac{N}{V} = \textit{const} = n$ and $\frac{E_{tot}}{N} = \textit{const} = k T$, the gas'es number of degrees of freedom grows to infinity and the surface degrees of freedom cease to matter. The gas pressure transfers to the usual $P_{gas} = n k T$ with $\frac{E_{tot}}{(N+1)} \rightarrow kT$. The obtained distribution of energy of gas particles Eq.\ref{eqstat-eq15} transfers to the usual Boltzmann distribution $\underset{N \rightarrow \infty}{\lim} N \frac{(E_{tot}-E)^{N-1}}{E_{tot}^N} = \frac{1}{k T} e^{-\frac{E}{k T}}$, with the same limit parameter $\underset{N \rightarrow \infty}{\lim} \frac{E_{tot}}{N} = k T$. Thus, the obtained for a finite number of particles expressions for the gas pressure and the particle energy distribution, in the thermodynamic limit transfer to usual ideal gas ones. In this way it could be traced that the same quantity appears in both expressions as temperature.

\section{Surface with multiple degrees of freedom}

The above results were obtained in a most simple case of a surface with only one mobile element. Let us now extend them to the surface of $S$ equivalent mobile elements, as shown at Fig.\ref{eqstat-fig1}(b). The treatment of this case and obtained results have no principal differences, only the complexity of expressions and calculations is increased. To obtain the distribution of coordinates of a surface element, we have first integrated the distribution Eq.\ref{eqstat-eq7} in a way described above, over all momenta components and particles coordinates, and received:

\begin{equation}\label{eqstat-eq32}
p(X_1) = const \!\!\!\!\!\! \underset{X_i=-X_{i max}}{\stackrel{X_{i max}}{\int \! \int}} \!\!\!\!\! (S L-\underset{i=1}{\stackrel{S}\sum} X_i)^N (\frac{2 E_{tot}}{k}-\underset{i=1}{\stackrel{S}\sum} X_i^2)^{N+\frac{S}{2}-1} \; dX_2 .. dX_S
\end{equation}

where

\begin{equation}\label{eqstat-eq32-1}
X_{i max} = \sqrt{\frac{2 E_{tot}}{k}-\underset{j=1}{\stackrel{i-1}\sum} X^2_j}, \quad i=2..S
\end{equation}

Further calculation of this integral meets certain mathematical difficulties. However, we managed to obtain the following remarkable mathematical relation for surface coordinates, analogous to the relation Eq.\ref{eqstat-eq10} for momenta:

\begin{equation}\label{eqstat-append-eq2}
\begin{array}{l}
\qquad \qquad \quad \underset{x=-\sqrt{b}}{\stackrel{\sqrt{b}}\int} (a-x)^N (b-x^2)^{N+S-1} \, _2F_1(\{ \frac{1-N}{2}, -\frac{N}{2} \},N+S\quad \;\;\,, (S-1)\frac{b-x^2}{(a-x)^2}) dx = \\
= \frac{\sqrt{\pi} \Gamma(N+S+1) }{(N+S) \Gamma(N+S+\frac{1}{2})} \;\;\; a^N \qquad \;\; b^{N+S-\frac{1}{2}} \qquad \;\; _2F_1(\{ \frac{1-N}{2}, -\frac{N}{2} \},N+S+\frac{1}{2}, \: S \qquad \;\: \frac{b}{a^2}) \\
\end{array}
\end{equation}

As can be seen, the integrand and the result of integration have the same structure. The procedure of calculation of this definite integral is simply in removal of the variable of integration from the integrand, with an increment of some integrand parameters. By taking as $a$ and $b$ in Eq.\ref{eqstat-append-eq2} the expressions of the form $S L-\underset{i}{\sum} X_i$ and $\frac{2 E_{tot}}{k}-\underset{i}{\sum} X_i^2$ and splitting off variables one by one, it is possible to consistently perform integration until the integration variables run out and the integral become calculated. As a result, after normalization, we have obtained the following distribution of coordinates of the surface elements:

\begin{equation}\label{eqstat-append-eq3}
\begin{array}{l}
p(X_1)=C \: (S L-X_1)^N (E_{tot}-\frac{k X_1^2}{2})^{N+S-\frac{3}{2}} \, _2F_1(\{ \frac{1-N}{2}, -\frac{N}{2} \},N+S-\frac{1}{2}, (S-1) \frac{2 E_{tot}-k X_1^2}{k (S L-X_1)^2});\\
\quad \;\; \: C=\frac{\sqrt{k} \, \Gamma(N+S) }{\sqrt{2\pi} \, \Gamma(N+S-\frac{1}{2}) \, (S L)^N E_{tot}^{N+S-1} \, _2F_1(\{ \frac{1-N}{2}, -\frac{N}{2} \},N+S, \frac{1}{S}\frac{2 E_{tot}}{k L^2})}\\
\end{array}
\end{equation}

Or, in another notation:

\begin{equation}\label{eqstat-append-eq4}
\begin{array}{l}
p(X_1)=C \: (1-\frac{X_1}{S L})^N (1-\frac{X_1^2}{A^2})^{N+S-\frac{3}{2}} \, _2F_1(\{ \frac{1-N}{2}, -\frac{N}{2} \},N+S-\frac{1}{2}, (S-1)\frac{A^2-X_1^2}{(S L-X_1)^2}) ; \\
\\
\quad \;\; \: C=\frac{\Gamma(N+S)}{A\sqrt{\pi} \, \Gamma(N+S-\frac{1}{2}) \, _2F_1(\{ \frac{1-N}{2}, -\frac{N}{2} \},N+S, \frac{1}{S}\frac{A^2}{L^2})}\\
\end{array}
\end{equation}

All the surface elements share this distribution. Let us remind, that it is valid only at $A<L$. For $A>L$, the distribution itself and its normalization constant will be different. The average displacement of the surface element $\overline{X_1}=\underset{X_1=-A}{\stackrel{A}\int} X_1 p(X_1) dX_1$, obtained from the distribution $Eq.\ref{eqstat-append-eq3}$, corresponds to the gas pressure:

\begin{equation}\label{eqstat-append-eq5}
P_{gas} = -\frac{k S \overline{X_1}}{H} = \frac{E_{tot}}{H L} \frac{N}{N+S} \frac{_2F_1(\{ \frac{1-N}{2}, -\frac{N}{2}+1 \},N+S+1, \frac{1}{S}\frac{2 E_{tot}}{k L^2})}{_2F_1(\{ \frac{1-N}{2}, -\frac{N}{2} \},N+S, \frac{1}{S}\frac{2 E_{tot}}{k L^2})}
\end{equation}

This is the exact equation of state of an ideal gas in the considered container with $S$ surface degrees of freedom.

For the average potential and kinetic energies per degree of freedom we have obtained:

\begin{equation}\label{eqstat-append-eq6}
\begin{array}{l}
\overline{\mathcal{E}}_P = \frac{E_{tot}}{2(N+S)} \frac{_3F_2(\{1+\frac{S}{2},\frac{1-N}{2},-\frac{N}{2}\},\{\frac{S}{2},N+S+1\},\frac{1}{S}\frac{2 E_{tot}}{k L^2} )}{_2F_1(\{\frac{1-N}{2},-\frac{N}{2}\},N+S,\frac{1}{S}\frac{2 E_{tot}}{k L^2})}\\
\overline{\mathcal{E}}_{K}=\overline{\mathcal{E}}_{W K}=\frac{\overline{\mathcal{E}}_{P K}}{2}=\frac{E_{tot}}{2N+S} \left(1- \frac{S}{2(N+S)}  \frac{_3F_2(\{1+\frac{S}{2},\frac{1-N}{2},-\frac{N}{2}\},\{\frac{S}{2},N+S+1\},\frac{1}{S}\frac{2 E_{tot}}{k L^2} )}{_2F_1(\{\frac{1-N}{2},-\frac{N}{2}\},N+S,\frac{1}{S}\frac{2 E_{tot}}{k L^2})} \right)

\end{array}
\end{equation}

Same as above, the average potential energy of the surface is greater than the kinetic energy per a degree of freedom. The kinetic energy is distributed uniformly.

We will not further proceed to the stationary walls limit $k \rightarrow \infty$ and then thermodynamic limit $N,V,E_{tot} \rightarrow \infty$, since they are similar to the case of $S=1$. If the wall's stiffness $k$ is not set infinite, but is very high, we can take $\frac{1}{k}$ as a small value ($\frac{A}{L}<<1$ or $\frac{1}{k} << \frac{L^2}{E_{tot}}$) and take series of the obtained above exact expressions:

\begin{equation}\label{eqstat-eq30}
\begin{array}{l}
P_{gas} \approx \frac{E_{tot} N}{H L (N+S)}(1-\frac{(N-1)(3N+2S)}{2S(N+S)(N+S+1)}\frac{E_{tot}}{L^2}\frac{1}{k}) \\
\\
\overline{\mathcal{E}}_K \approx \frac{E_{tot}}{2(N+S)}(1 - \frac{N(N-1)}{2 S (N+S) (N+S+1)}\frac{E_{tot}}{L^2}\frac{1}{k})\\
\\
\overline{\mathcal{E}}_P-\overline{\mathcal{E}}_K \approx \frac{E_{tot}N(N-1)}{2 S^2 (N+S)(N+S+1)}\frac{E_{tot}}{L^2}\frac{1}{k} \\
\\
\overline{V} = H (L-\overline{X}) \approx H L + \frac{N H E_{tot}}{L S (N+S)}\frac{1}{k}\\
\end{array}
\end{equation}

From where, leaving the first order terms, we obtain:

\begin{equation}\label{eqstat-eq31}
P_{gas} \overline{V} \approx 2 N \overline{\mathcal{E}}_K + \frac{2 S (2N+S)}{(N+S)(N-1)} (\overline{\mathcal{E}}_P - \overline{\mathcal{E}}_K)
\end{equation}

The second term here is a small amendment to the usual equation of state. It appears, if the vessel walls are not perfectly rigid and there is a small influence of the surface. It's interesting to note that $\overline{\mathcal{E}}_P-\overline{\mathcal{E}}_K \approx \overline{\mathcal{E}}_K \frac{N(N-1)}{S^2 (N+S+1)}\frac{E_{tot}}{L^2}\frac{1}{k}$ may not be small compared to $\overline{\mathcal{E}}_K$, if $N >> S^2$.

Thus, if the container walls are not perfectly stiff and perform small motion, the gas pressure is not exactly equal to $P_{gas} = 2 n \overline{\mathcal{E}}_K$. The pressure level will be higher due to the influence of the surface degrees of freedom.

\section{Discussion}

The discussed equipartition violation and the appearance of the surface energy is a consequence of the fact that the particle and the surface element coordinates are not independent, but bound with inequalities. The internal particles should be located inside the varying surface. Due to the internal pressure, the container is stretched from volume $V$ to $\overline{V}$. That leads to the appearance of the excessive potential energy of the surface, which we here theoretically introduce and calculate. Such an energy appears when the surface prevents internal particles to scatter due to their thermal motion. But this is not the only one reason for the surface energy to appear. Generally, the main part of the energy of a condensed matter interface is related to a different numbers of nearest neighbours for bulk and surface atoms, i.e. internal surface structure. Nevertheless, the system will still tend to reduce its surface energy, regardless of its nature. If it is related to the reduction of system's volume, there will be interplay of the surface energy and internal pressure. It is dependent on the geometry and other details. In this paper, we have considered one of the type and the provided results are valid when the considered model applies. It is not always the case. But other geometries and surface models may be considered in a similar way.

\section{Conclusion}

We have considered a gas of a finite number of particles in a finite-sized vessel with surface degrees of freedom. A complete statistical description of the system was constructed on a pre thermodynamic limit basis, which can be traced up to the laws of Hamiltonian mechanics. The paper provides distributions of coordinates and kinetic energies of the gas particles and the surface elements. The distribution of surface coordinates allows determination of the mean spring compression, from which the gas pressure can be directly obtained. In the limit of an infinite spring stiffness, the vessel walls became stationary. In this way, we have derived the previously unknown equation of state for a gas of a finite number of particles. It differs from the usual equation of state in the change of $\frac{1}{2}kT$ to $\overline{\mathcal{E}}_K$ - the mean kinetic energy per a degree of freedom. The consequent thermodynamic limit passage allows to trace how a temperature appears instead of mechanical quantities. Thus, the paper contains the derivation of the usual ideal gas equation of state and shows theoretically that $k T$ there is the same limit value as in the particle energy distribution. Without the limit passages, the equation of state depends on the number of the surface degrees of freedom, its geometry and parameters. We have obtained it explicitly for the considered model of a finite sized system with finite numbers of bulk and surface degrees of freedom.

The energy of the system was found to be distributed unevenly. The mean potential energy of a surface element is higher than the mean kinetic energy per degree of freedom. Kinetic energy is evenly distributed, despite the energy distributions of particles and surface elements being very different. The lack of energy equipartition is a consequence of the fact that the phase variables are not independent. Because of that, the hypersurface of constant energy is not entirely available to the trajectory, as is assumed in the proofs of the equipartition theorem. This lack of equipartition opens a fundamental way to theoretically introduce the excessive surface energy and study the interplay between the system's surface and internal pressure.

\end{document}